\documentclass[14pt,a4paper]{atlasnote}
\graphicspath{{figures/}}
\usepackage{atlasphysics}
\usepackage{subfigure}
\usepackage{mathrsfs}

\title{SM Higgs boson searches in the early ATLAS data}
\author{I. Tsukerman for the ATLAS Collaboration}
\draftversion{1.0}
\abstracttext{
ATLAS exclusion and discovery potentials of Standard Model Higgs boson searches at the LHC at 14 TeV, 10 TeV and 
7 TeV center-of-mass energy are reviewed. For a LHC center-of-mass energy of 7 TeV and integrated luminosity of 1 fb$^{-1}$,
contributions from the important decay channels $\Hboson \rightarrow \Wboson\Wboson^{(*)}
\rightarrow \ell\nu\ell\nu$, $\Hboson \rightarrow \Zboson\Zboson^{(*)} \rightarrow \ell\ell\ell\ell$ and 
$\Hboson \rightarrow \gamma\gamma$ are considered, based 
on recent full Monte Carlo simulations at 14 and 10 TeV and the 
cross-section rescaling for the 7 TeV center-of-mass energy.
First measurements of backgrounds to Standard Model Higgs boson search are also presented.
}

\begin{document}
\clearpage
\section{Introduction}

The primary objective of the Large Hadron Collider (LHC) 
is to study the origin of electroweak symmetry breaking.
Within the Standard Model (SM), the Higgs mechanism \cite{higgs1}\cite{higgs2}\cite{higgs3} is invoked
to explain this breaking, but the predicted SM Higgs boson still remains undiscovered.
The direct search at the $e^{+}e^{-}$ collider LEP has led to the lower
limit on its mass of 114.4 GeV at 95\% confidence level (C.L.) \cite{lep}.
In addition, high precision electroweak data constrain the Higgs boson mass 
via their sensitivity to loop corrections. The corresponding upper limit 
is $\mH\leq$ 185 GeV at 95\% C.L., provided that the result of the direct search at LEP is also used in the
determination of this limit \cite{lepewwg}. Combined analysis of
the Tevatron data from CDF and D0, based on the integrated luminosity 
up to 6.7 fb$^{-1}$ at 
1.8 TeV, yields a 95\% C.L. exclusion of a Higgs boson with mass between 
158 GeV $\leq \mH \leq$ 175 GeV \cite{fnal}.  

The ATLAS experiment \cite{atlas} at the LHC is designed to search for the Higgs boson over a
wide mass range. Since the end of March 2010, LHC was running at a total energy of 7 TeV
for pp-collisions and by the end of the year reached the integrated luminosity
of less than 50 pb$^{-1}$.
In this report, the ATLAS sensitivity 
to discover or exclude the SM Higgs boson and recent developments
to enhance this sensitivity are summarized. In addition, the results of first
measurements of backgrounds to SM Higgs boson production are presented.   
\section{SM Higgs search channels}
\subsection{SM Higgs production and decays at the LHC}
The dominant production mechanism  for a SM Higgs boson production
is gluon fusion ($gg\rightarrow~\Hboson$). At 14 TeV, it has a cross-section at NLO of 20--40 pb \cite{atl07}
in the Higgs boson mass range between 114 
and 185 GeV.\footnote{Although the NNLO corrections increase the cross-section by a factor of $\approx$1.5, ATLAS
used conservative (NLO) approach in the sensitivity studies.}
The vector-boson fusion (VBF) process ($qq\rightarrow qq\Hboson$) has a factor of eight smaller 
NLO cross-section than the $gg\rightarrow~\Hboson$ mode but has a characteristic signature: 
the Higgs boson is accompanied by two energetic, mainly forward-going, jets. 

The SM Higgs boson is predicted to have many decay channels 
with branching ratios strongly dependent on its mass.
The evaluation of the search sensitivity
of various channels should take into account the cross-sections
of the relevant backgrounds.
At low Higgs boson mass the dominant decay mode is through $b\bar{b}$. However,
due to the enormous QCD backgrounds this channel is not optimal for the SM Higgs discovery.
The $\Hboson \rightarrow \Wboson\Wboson^{(*)}
\rightarrow \ell\nu\ell\nu$ final state is the main search channel in a wide $\mH$ range
below 200 GeV, due to a large branching ratio. For the 
$\Hboson \rightarrow \Zboson\Zboson^{(*)} \rightarrow \ell\ell\ell\ell$ mode,
the resulting decay probability is small but the signal is easy to trigger on 
and allows for a full and precise reconstruction of the Higgs mass.
The $\gamma\gamma$ final state has a small branching fraction but a very
clean signature. 
ATLAS and CMS Collaborations have performed 
extensive GEANT-based Monte Carlo (MC) \cite{gean1} studies
with full simulation and reconstruction at 14 TeV \cite{atlcsc},\cite{cmstdr} to 
explore the feasibility of observing different Higgs decay channels. It was concluded that 
with an integrated LHC luminosity of 2 fb$^{-1}$
ATLAS alone is able to discover the SM Higgs boson in the mass range between 143 and 179 GeV. 
ATLAS Collaboration also estimated 
$\Hboson\rightarrow \Wboson \Wboson^{(*)} \rightarrow \ell\nu\ell\nu$
exclusion potential at 10 TeV and 200 pb$^{-1}$ \cite{atl10}. When it became clear
that the LHC will work at 7 TeV during a prolonged period of time, the related study was performed 
for this energy \cite{atl07}, by
re-scaling the expectations from detailed analyses at 10 or 14 TeV and  
using re-weighted cross-section ratios or parton distribution functions.\footnote{Note, that the signal and background
cross-sections at 7 TeV are roughly four times smaller than at 14 TeV.}  
Similar cross section rescaling is performed also for other Higgs decay channels [8].
\subsection{$\Hboson\rightarrow \Wboson \Wboson^{(*)} \rightarrow \ell\nu\ell\nu$}
\begin{figure*}[t]
\centering
\includegraphics[width=60mm]{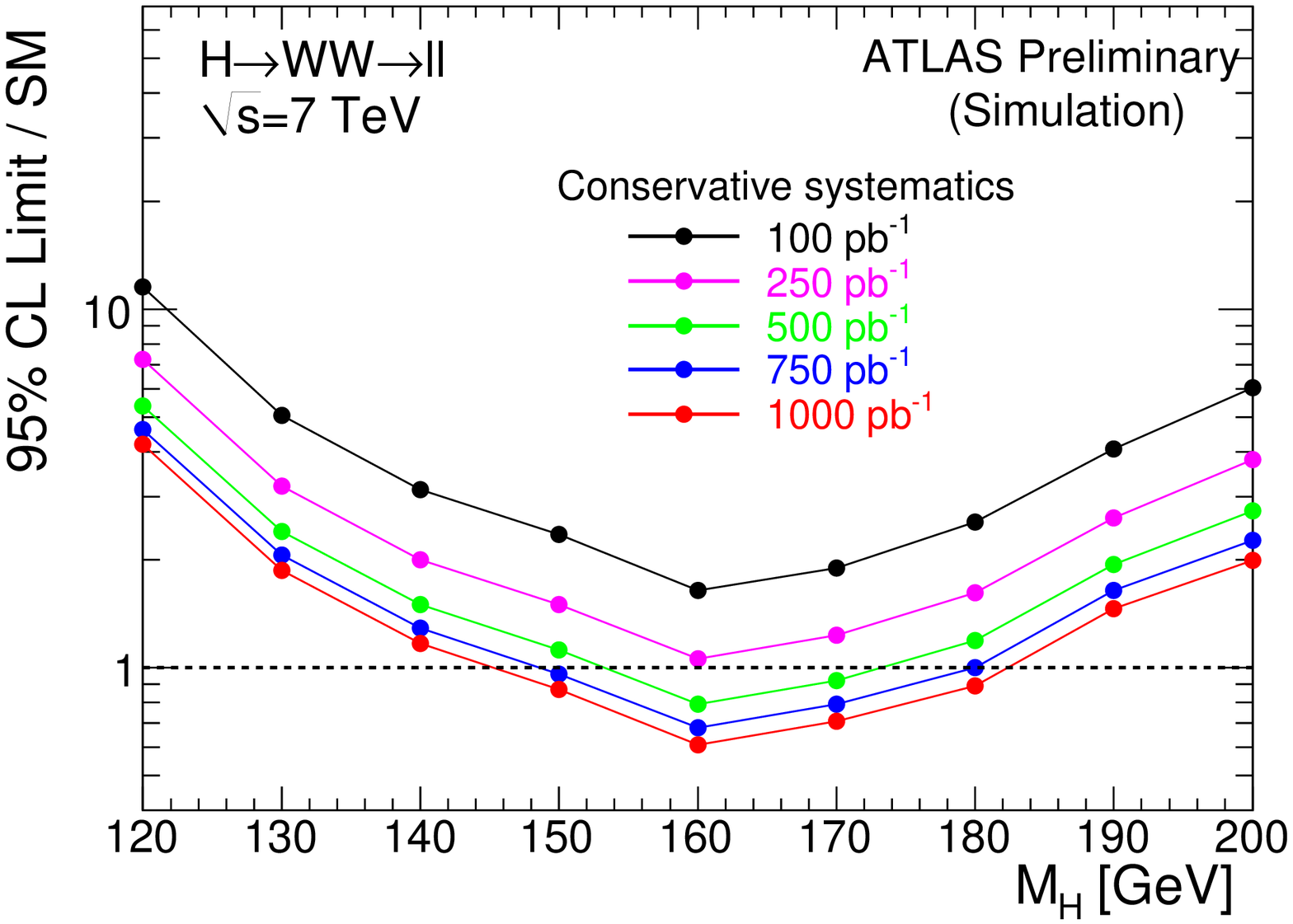}
\includegraphics[width=60mm]{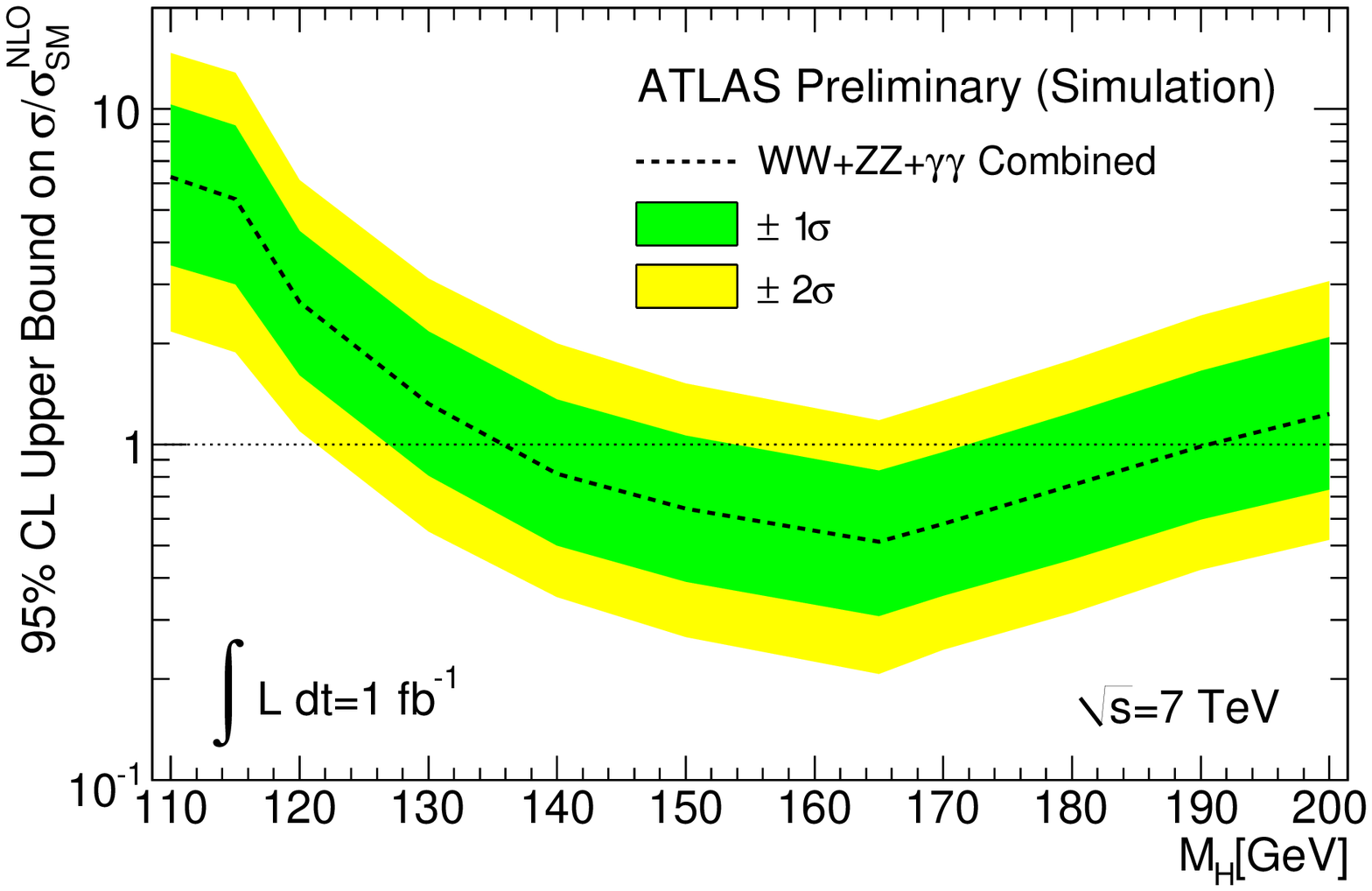}
\caption {
Left: expected 95\% C.L. upper limits on the SM Higgs boson production rate normalized to the SM prediction, 
for different integrated luminosity scenarios 
at 7 TeV, for the $\Hboson\rightarrow \Wboson \Wboson^{(*)} \rightarrow \ell\nu\ell\nu$ channel alone.
Right: the expected upper bound on the Higgs boson production cross-section after collecting 1 fb$^{-1}$ 
of integrated luminosity at 7 TeV in ATLAS. $\Hboson \rightarrow \Wboson\Wboson^{(*)}
\rightarrow \ell\nu\ell\nu$, $\Hboson \rightarrow \Zboson\Zboson^{(*)} \rightarrow \ell\ell\ell\ell$ and
$\Hboson\rightarrow \gamma\gamma$ decay channels
are combined.
The coloured area represent the 1$\sigma$ and 2$\sigma$ error bands.} 
\label{lumiexcl}
\end{figure*}

At low integrated luminosity, the decay channel $\Hboson\rightarrow \Wboson \Wboson^{(*)}$ provides the best sensitivity for
the SM Higgs boson search in ATLAS. 
However, unlike with the $\Hboson\rightarrow \Zboson\Zboson^{(*)} \rightarrow 4\ell$ and $\Hboson\rightarrow \gamma\gamma$ channels, 
a full mass reconstruction is impossible in this channel, therefore an
accurate background estimate is very important. 
The main background is $q\bar{q},gg\rightarrow \Wboson\Wboson^{(*)}$-production giving two oppositely charged leptons in the final state. 
Additional smaller background contributions are $t\bar{t}$ and $\Wboson t$. These processes
can be suppressed, in particular, by exploiting the spin correlation between the two final state leptons,
and applying an upper cut on the lepton pair invariant-mass.
Another important background comes from $\Wln$ + jets, with one of the jets misidentified as a lepton.
Data-driven methods have been developed to estimate this background.\footnote{To estimate a background contribution in a signal region,
one selects the background-dominated control region and uses the MC prediction to extrapolate this background contribution into the signal region.}
Fig.~\ref{lumiexcl} (left) shows the exclusion reach as a function of the Higgs boson mass $\mH$ for different integrated luminosity 
scenarios at 7 TeV. With 1 fb$^{-1}$, ATLAS expects to
exclude a region 140 GeV $\leq \mH \leq$ 185 GeV at 95\% C.L., see Fig.~\ref{lumiexcl} (right).
A 5$\sigma$-discovery for the SM Higgs boson with $\mH$ around 160 GeV is expected in the $\Hboson\rightarrow \Wboson \Wboson^{(*)} \rightarrow \ell\nu\ell\nu$
mode alone with the integrated LHC luminosity of $\approx$5 fb$^{-1}$. 
\subsection{$\Hboson\rightarrow \Zboson\Zboson^{(*)} \rightarrow 4\ell$}
The ``golden'' channels (4$\mu$, 2$e$2$\mu$ and 4$e$ final states of $\Zboson\Zboson^{(*)}$ decays) 
are expected to have a good discovery potential in a wide mass range 
(except $\mH\leq$130 GeV and $\mH\approx 2\mW$). 
The irreducible background is a non-resonant $ZZ^{(*)}$-continuum. For a low mass
region with $\mH \leq$ 180 GeV, there are additional reducible backgrounds 
mainly from $\Zboson b\bar{b}$, $t\bar{t}$, $\Wboson\Zboson$ and $\Zboson+jet$ processes.
They can be suppressed by using the impact parameter and lepton isolation requirements. 
Simulations of the signal and the background processes  made at the NLO level
show that at 7 TeV energy and 1 fb$^{-1}$ integrated luminosity a 95\% C.L. exclusion 
is not expected. 
\subsection{$\Hboson\rightarrow \gamma\gamma$}
Despite of a small (0.2\%) branching ratio in the Higgs mass region from 120--140 GeV,
$\Hboson\rightarrow \gamma\gamma$ remains a promising channel due to a very clean signal signature 
is very clean. Irreducible backgrounds originate from the continuum, 
$q\bar{q}, gg\rightarrow \gamma\gamma$. Reducible backgrounds are mostly
due to $\gamma$-jet and jet-jet events with one or more jets misidentified
as photons. The high-granularity electromagnetic calorimeter of ATLAS
is capable of determining the photon direction with 
a great precision and has an excellent energy and angular resolution. It 
permits the suppression of the reducible backgrounds well below the $\gamma\gamma$-continuum.
Like in the case of $\Hboson\rightarrow \Zboson\Zboson^{(*)} \rightarrow 4\ell$,
with 7 TeV and 1 fb$^{-1}$ one cannot exclude at 95\% C.L. the SM Higgs boson with the 
$\Hboson\rightarrow \gamma\gamma$ channel alone. 
\subsection{Summary of the SM Higgs exclusion potential}
Figure~\ref{lumiexcl} (right) shows a combined exclusion reach
of the inclusive Higgs boson search with the decay modes 
$\Hboson\rightarrow \Wboson\Wboson^{(*)}\rightarrow \ell\nu\ell\nu$, 
$\Hboson\rightarrow \Zboson\Zboson^{(*)} \rightarrow 4\ell$ and
$\Hboson\rightarrow \gamma\gamma$ in ATLAS. In this plot,
conservative systematic errors for 
$\Hboson\rightarrow \Wboson \Wboson^{(*)} \rightarrow \ell\nu\ell\nu$ mode,
conservative NLO calculation of the signal cross-sections and 10\% uncertainty on the LHC integrated luminosity are assumed.
With 7 TeV LHC energy and 1 fb$^{-1}$ of integrated luminosity it would be possible to
exclude the region 135 GeV$\leq \mH \leq$ 188 GeV at 95\% C.L.
The exclusion limits could be improved by adding channels such as VBF $\Hboson\rightarrow\tau\tau$
and $V\Hboson \rightarrow b\bar{b}$ at low $\mH$, as well as
$\Hboson\rightarrow \Zboson\Zboson^{(*)} \rightarrow \ell\ell\nu\nu$ and 
$\Hboson\rightarrow \Zboson\Zboson^{(*)} \rightarrow \ell\ell b\bar{b}$ decays
at high $\mH$ values. Furthermore, using NNLO signal cross-sections, instead of NLO,
will make the exclusion region significantly wider. Finally, one could benefit from possible LHC running at 8 
or 9 TeV in 2011, instead of 7 TeV. The results based on all these improvements
are summarized in a recent Ref.\cite{atl09}.
With 2 fb$^{-1}$, 8 TeV and the analysis improvements, the 95\% C.L. exclusion range will span from 114 to over 500 GeV.
\section{First ATLAS measurements of backgrounds to the SM Higgs searches}
Based on 310 nb$^{-1}$ data at 7 TeV, ATLAS observed $\Wboson+jet$ background to
the $\Hboson\rightarrow \Wboson \Wboson^{(*)} \rightarrow \ell\nu\ell\nu$ search \cite{wwbgr}
and made data-driven estimation of the dijet and $\Wboson+jet$ backgrounds for the $\Hboson\rightarrow\tau\tau \rightarrow \ell h$
search \cite{taubgr}. 
\subsection{Observation of the $\Wboson+jet$ background to the 
$\Hboson\rightarrow \Wboson \Wboson^{(*)} \rightarrow \ell\nu\ell\nu$ search}
\begin{figure*}[t]
\centering
\includegraphics[width=50mm]{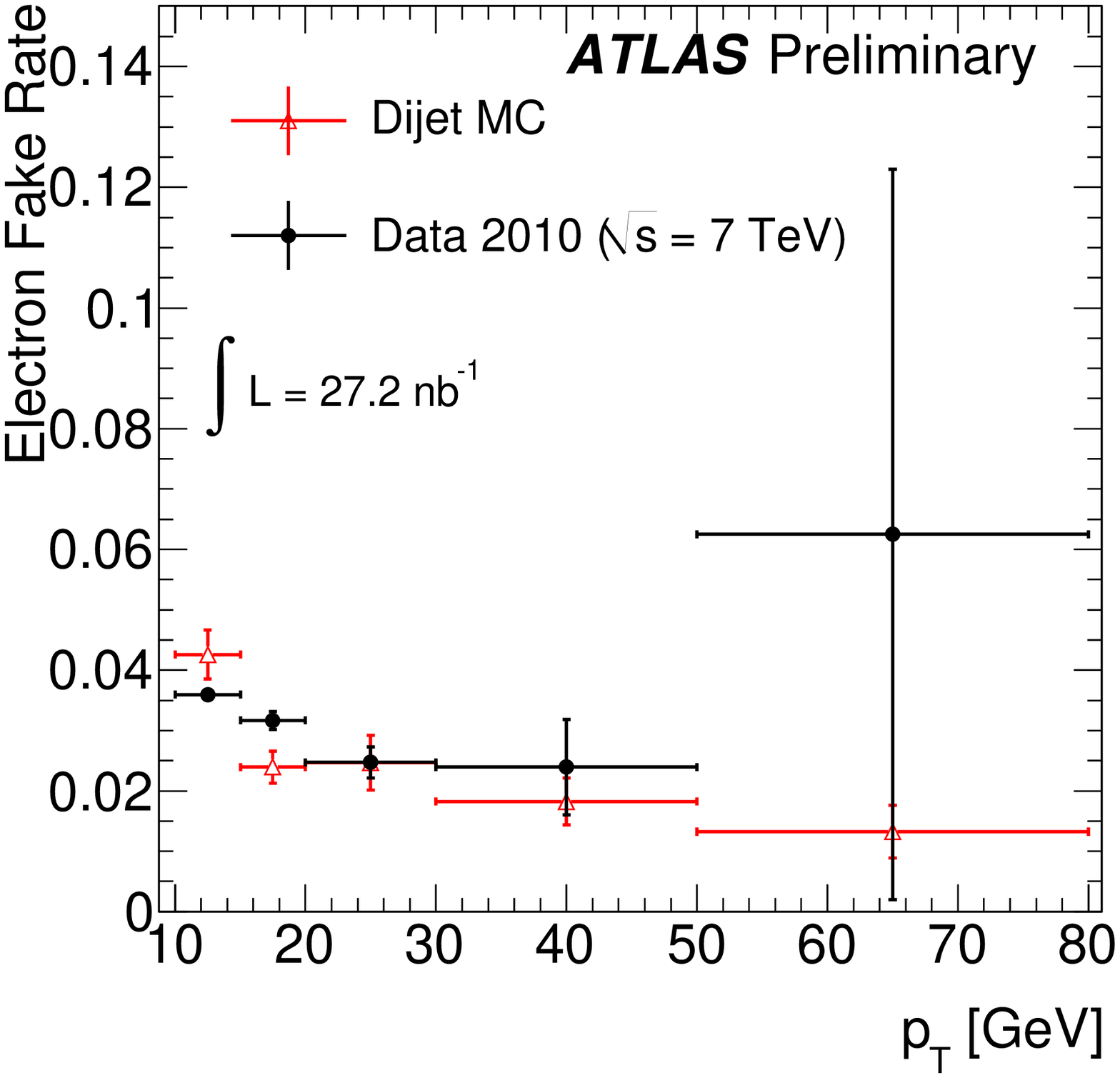}
\includegraphics[width=50mm]{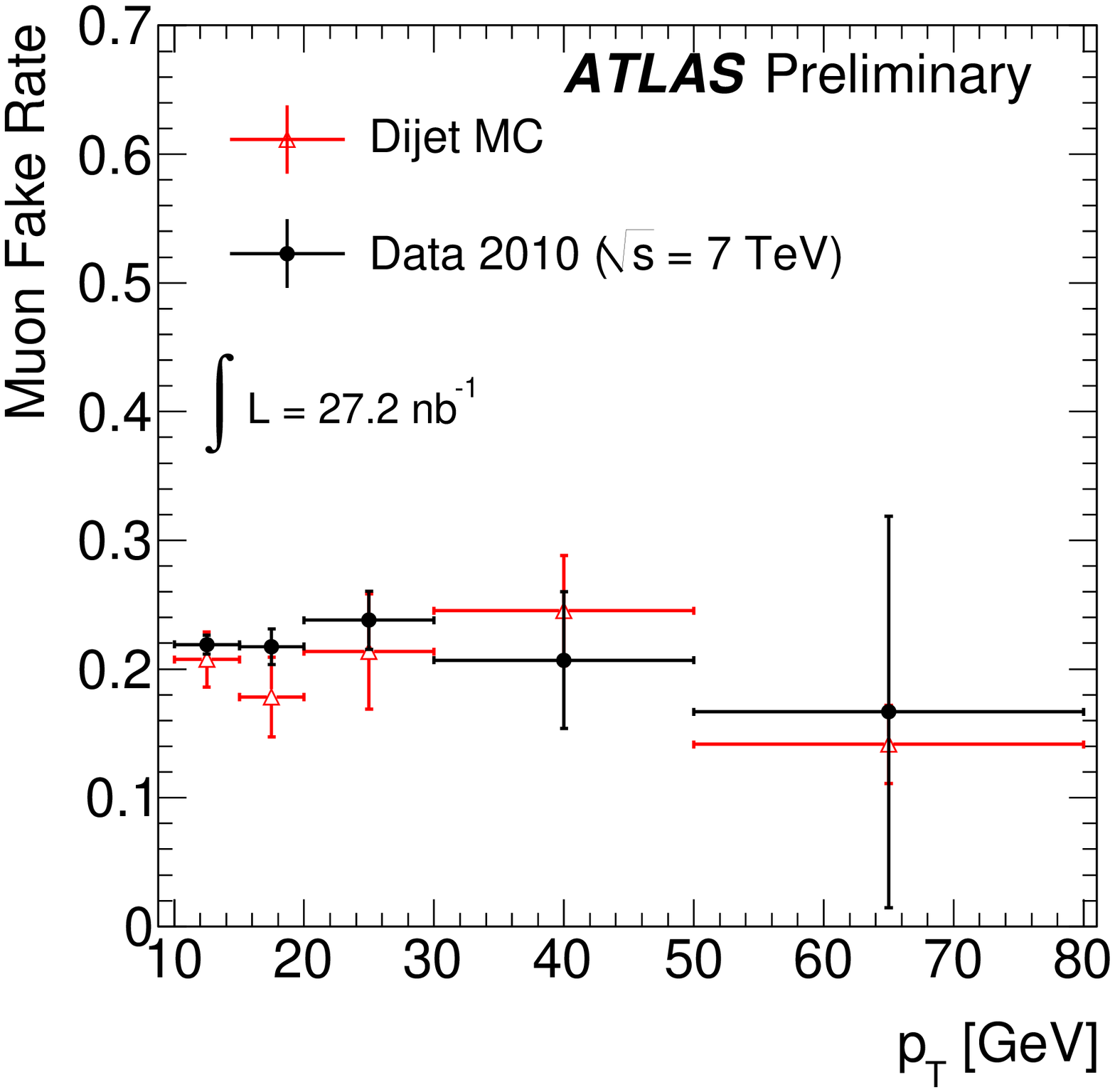}
\caption{
Fake lepton rates as function of lepton $\pT$. The fake rate measured from dijet data is compared with the
MC prediction. 
Left: electrons. Right: muons. }
\label{fake}
\end{figure*}

Nine events with  large missing transverse momentum and two lepton candidates, one of which passes the tighter and 
the other one the looser identification criteria \cite{atlas}, were observed. This is consistent with the MC
expectation. It therefore validates our MC based sensitivity studies. This data sample is dominated by $\Wboson+jet$ events and will be used as
a control region to estimate the $\Wboson+jet$ background contribution from the signal-like region defined for
the Higgs boson search, as demonstrated in Ref.\cite{atl10}. In addition to a pure
MC estimate, the contamination from the QCD background in this control region is also estimated by means of data-driven methods.
The estimation of the $\Wboson+jet$ contribution to the Higgs boson search requires the knowledge of the
rate of fake lepton events in which a loosely identified lepton candidate is also identified using tighter identification criteria.
Fake lepton rates down to the lepton transverse momenta of 10 GeV are obtained from
dijet and $\gamma$ + jet data samples and agree well with the MC predictions. 
They are shown in Fig.~\ref{fake}, separately for electrons and muons. 
\subsection{Data-driven background estimation for the $\Hboson\rightarrow\tau\tau \rightarrow \ell h$ search}
To estimate the background in this channel, events with an electron or a muon having, in addition, a hadronically decaying tau lepton
of the opposite charge and a large missing transverse energy, were selected. Candidate events
with $\MET \geq$ 20 GeV were found in both electron and muon channels -- 12 and 17 events, respectively.
This is consistent with the total of 25$\pm$9 events expected from a data-driven background estimation based on the control regions 
with low missing transverse energy and with the same sign charge of the lepton and hadronically decaying tau lepton.. The observed shape of the visible mass
distribution also agrees with the estimation. With an additional requirement of a large transverse
mass, 3 and 7 candidate events remain in the electron and muon channels, respectively. They are expected
to come mainly from the $\Wboson+jet$ and $\Zboson+jet$ processes, representing backgrounds for the
 $\Hboson\rightarrow\tau\tau \rightarrow \ell h$ search. This study provides a starting point for future
Higgs boson searches in di-tau channels and the corresponding data-driven background estimation.
\section{Conclusion}
\noindent 
The ATLAS simulations show that at 7 TeV and with the integrated LHC luminosity of 1 fb$^{-1}$
the SM Higgs boson can be excluded at 95\% C.L. in the range 135 GeV $\leq \mH \leq$ 188 GeV,
by combining $\Hboson \rightarrow \Wboson\Wboson^{(*)}
\rightarrow \ell\nu\ell\nu$, $\Hboson \rightarrow \Zboson\Zboson^{(*)} \rightarrow \ell\ell\ell\ell$
and $\Hboson\rightarrow \gamma\gamma$ channels
and using conservative assumptions. Should 2 fb$^{-1}$ and 8 TeV 
be available, one can cover the exclusion range from 
114 to over 500 GeV by adding extra decay channels and improving the analysis. 
First measurements of backgrounds to the SM Higgs boson production 
have been performed. They will be extended as more data become available.

%
%
\section*{Acknowledgements}
We wish to thank CERN for the efficient commissioning and operation of the LHC during this initial high-energy 
data-taking period as well as the support staff from our institutions without whom ATLAS could not be operated efficiently.
We acknowledge the support of our all funding agencies. The crucial computing support from all WLCG partners is acknowledged gratefully.


\bibliographystyle{atlasnote}
\bibliography{iitiheplhc2010}

\end{document}